\documentclass[aps,prc,showpacs,onecolumn,preprintnumbers,superscriptaddress,nofootinbib,floatfix,10pt]{revtex4-1}
\usepackage{amsfonts,amssymb,stmaryrd,latexsym,amsmath,braket}
\usepackage{graphicx,subfigure}
\usepackage{comment}
\usepackage{times}
\usepackage{slashed}
\usepackage{bm}
\usepackage{braket}
\usepackage{chapterbib}
\usepackage{color}
\usepackage{longtable}
\usepackage{hyperref}
\hypersetup{colorlinks=true,linkcolor=blue,citecolor=blue,urlcolor=blue}
\usepackage{natbib}
\usepackage{bibunits}
\usepackage{epstopdf}
\makeatletter
\let\saved@includegraphics\includegraphics
\AtBeginDocument{\let\includegraphics\saved@includegraphics}
\makeatother
\usepackage{booktabs}
\graphicspath{{figures/}}
\newcommand{\be}{\begin{equation}}
\newcommand{\ee}{\end{equation}}
\begin{document}
\begin{bibunit}[apsrev4-2]
\title{Investigation of the $K^{-}pp$ Bound State via the 
\( K^{-} + {}^{3}\mathrm{He} \) Reaction}
\author{Sajjad Marri}
\email{marri@iut.ac.ir}
\affiliation{Department of Physics, Isfahan University 
of Technology, Isfahan 84156-83111, Iran}
\author{Ahmad Naderi Beni}
\affiliation{Department of Basic Sciences, Technical and 
Vocational University (TVU), Tehran, Iran}
\begin{abstract}
Using the four-body Alt-Grassberger-Sandhas (AGS) equations for the \( K^{-}ppn \) 
system, we investigate the possible formation of a \( K^{-}pp \) quasi-bound state 
through the low-energy \( K^{-} + {}^{3}\mathrm{He} \) reaction. The neutron missing 
mass spectrum in the final state was calculated for different models of the 
\( \bar{K}N \) interaction. The results indicate that, irrespective of the specific 
nature of the \( \Lambda(1405) \) structure or the details of the \( \bar{K}N \) 
interaction model, a signal corresponding to the \( K^{-}pp \) quasi-bound state 
could appear in the \( \pi \Sigma p \) mass spectrum. This supports the feasibility 
of observing the \( K^{-}pp \) cluster in low-energy kaon-induced reactions on helium-3.
\end{abstract}
\pacs{13.75.Jz, 14.20.Pt, 21.85.+d, 25.80.Nv}
\maketitle
\section{Introduction}
\label{intro}
The interaction between antikaons (\(\bar{K}\)) and nucleons plays a 
significant role in the study of strongly interacting kaonic systems. 
This interaction, characterized by its attractive nature in the 
isospin-zero channel, has garnered considerable attention due to its 
connection to exotic nuclear states and its implications for understanding 
low-energy quantum chromodynamics (QCD) \cite{l1,l2,l3,l4,l5}. One of 
the most intriguing manifestations of this interaction is the possible 
existence of the \(K^-pp\) system, a three-body configuration composed 
of an antikaon and two protons \cite{k1,k2,k3,k4,k5,k6,k7,k8,k9,k10,k11,k12,k13}. 
The study of this system provides insight into the underlying meson-baryon 
dynamics and the nature of the \(\Lambda(1405)\) resonance, which is 
dynamically generated by the coupling of \(\bar{K}N\) and \(\pi\Sigma\) 
channels~\cite{l1,l2}.

The theoretical prediction of deeply bound antikaonic nuclear states 
was first proposed by Akaishi and Yamazaki \cite{k1,k2}, who suggested 
that the strong attraction of the antikaon in the \(I=0\) channel could 
lead to the formation of compact, high-density nuclear configurations. 
Their calculations indicated the existence of a bound \(K^-pp\) state 
with a binding energy of approximately 48 MeV and a width of about 61 
MeV. Subsequent studies employing different theoretical approaches, 
including Faddeev calculations \cite{k3,k4,k5,k6}, have investigated 
the \(K^-pp\) system with varying predictions for its binding energy and 
decay width. Recent coupled-channel three-body studies have shown that 
explicit particle-channel dynamics can significantly modify the properties 
of the $K^-pp$ quasi-bound state \cite{sh25}. These discrepancies highlight 
the complexity of the system and the challenges in reliably describing the 
interplay between the antikaon and the two-nucleon core and the sensitivity 
of the results to the input interactions.

Experimental efforts to observe the \(K^-pp\) state have included 
proton-proton collisions, stopped-kaon reactions, and photoproduction 
experiments. The DISTO experiment at SATURNE reported a structure in 
\(pp\) collisions suggesting a binding energy of around 103 MeV and a 
width of about 118 MeV \cite{YAM20}. Similarly, the FINUDA collaboration 
observed a peak in the \(\Lambda p\) invariant mass spectrum consistent 
with a binding energy of roughly 115 MeV and a width of 67 MeV \cite{FIN5}. 
However, these findings remain controversial due to possible final-state 
interaction effects \cite{Do02,mag6}. In particular, the FINUDA results 
were later re-examined by the AMADEUS collaboration \cite{amd19}, which 
concluded that the observed spectrum can be fully explained by \(K^-\) 
multi-nucleon absorption processes, without requiring a \(K^- pp\) component. 
The J-PARC E27 experiment, investigating the reaction \(\pi^+ + d\), also 
reported a broad structure, though its interpretation was complicated by 
background processes \cite{E271,E272}. The CLAS collaboration at Jefferson 
Lab studied kaonic clusters through photoproduction, while the AMADEUS 
experiment at DA\(\mathrm{\Phi}\)NE investigated low-energy kaonic systems 
using low-momentum kaons \cite{ama9}. Despite these efforts, no conclusive 
evidence for the \(K^-pp\) state was found. The lack of a distinct signal 
may indicate a very broad width or the absence of a conventional three-body 
bound state. 

\begin{figure}[h]
\centering
\includegraphics[width=12.2cm]{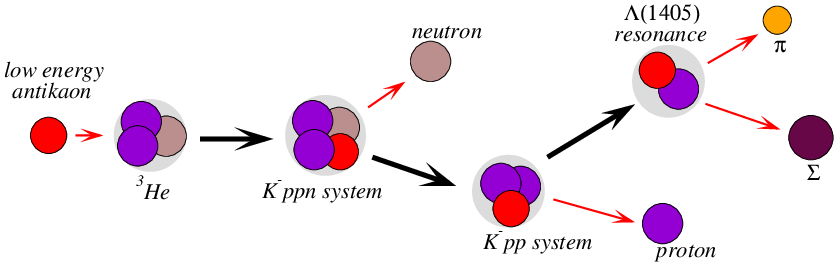} 
\caption{(Color online) Schematic representation of the transition from 
\( K^{-} + \mathrm{^{3}He} \) to \( n + (\pi\Sigma p) \) via the intermediate 
\( N + (\bar{K}NN)_{s=0} \) state.}
\label{fig0}
\end{figure}

The interaction between an antikaon and a helium-3 nucleus (\(K^- + ^3\text{He}\)) 
provides a promising avenue for probing the \(K^-pp\) state through kaon-induced 
reactions. The J-PARC E15 experiment utilized such interactions to study the system 
via neutron missing mass spectroscopy (Fig.~\ref{fig0}). Investigating the 
\(K^- + ^3\text{He}\) reaction is particularly valuable, as it may help isolate 
signals of the \(K^-pp\) state while reducing background contributions. Their 
results suggested a signal corresponding to a bound state with a binding energy of 
approximately 42 MeV and a decay width of 100 MeV \cite{E1524,E1520}. However, 
challenges persist: low-momentum kaon beams have low production rates, and low-energy 
neutrons are difficult to detect due to their weak interaction and low kinetic energy. 
Consequently, the J-PARC beamline is optimized for a kaon momentum of about 1~GeV/\(c\), 
balancing production efficiency, detection, and resolution. While low-energy studies 
may offer theoretical benefits, they remain experimentally demanding.

Sekihara and collaborators analyzed the \(K^- + {}^3\text{He}\) reaction to 
interpret the \(\Lambda p\) invariant mass spectrum reported by E15~\cite{Se16}. 
They considered scenarios where the incident kaon induces a reaction ejecting 
a fast neutron, leaving behind a residual two-nucleon system interacting with 
the antikaon. This intermediate system could either produce a non-resonant 
\(\Lambda(1405)\)-proton pair or form a quasi-bound \(\bar{K}NN\) state. Their 
simulations indicated that only the latter could reproduce the observed spectrum, 
with features attributed to both the decay of the quasi-bound state and kinematical 
enhancements from quasi-free scattering.

Previous theoretical studies of kaon-induced reactions on light nuclei, in 
particular the works by Ohnishi \textit{et al.} \cite{sh13} and Marri 
\textit{et al.} \cite{s19}, have investigated the $\bar{K}NN$–$\pi\Sigma N$ 
system using non-relativistic Faddeev calculations. Both works solved the 
three-body problem to explore the signature of the $K^-pp$ quasi-bound state. 
The main distinction lies in the interaction models: Ohnishi \textit{et al.} 
employed chiral $\bar{K}N$ potentials \cite{k10}, whereas Marri \textit{et al.} 
used phenomenological potentials developed by Shevchenko \cite{shev3,shev4}. 
Furthermore, Marri \textit{et al.} also observed contributions from the 
$\Lambda(1405)$ resonance in addition to the $K^-pp$ signal. In the present 
work, we extend these studies by performing a four-body calculation of the 
$K^- + {}^3\mathrm{He} \to \pi\Sigma p n$ reaction, which allows us to explicitly 
study the neutron missing-mass spectrum and provide absolute cross sections, 
thus offering more direct guidance for experimental investigations.

This study examines the feasibility and advantages of the reaction 
\(K^{-} + \,^{3}\mathrm{He} \rightarrow n + \pi\Sigma p\) as a probe for the 
\(K^-pp\) bound state. The hypothesized state is expected to decay into multiple 
final states, including \(\pi\Sigma p\). Since the \(\pi\Sigma\) subsystem is 
closely tied to the \(\Lambda(1405)\), often regarded as a \(\bar{K}N\) quasi-bound 
state, detecting \(\pi\Sigma\) pairs with a proton offers insight into the decay 
mechanisms of \(K^-pp\). Neutron detection enables missing mass reconstruction, 
providing a powerful tool for isolating signals from competing background processes. 
High-resolution invariant mass reconstruction of the \(\pi\Sigma p\) system, combined 
with neutron kinematics, enhances sensitivity to the binding energy and width of 
the state. The complexity of the multi-particle final state allows for improved 
discrimination between genuine bound states and final-state interaction effects, 
increasing the robustness of the experimental conclusions.

In this work, we apply the four-body Faddeev approach to the \(K^- + ^3\text{He}\) 
system to calculate the neutron missing mass spectrum and assess the presence of a 
\(K^-pp\) bound state. Our framework incorporates recent models of the \(\bar{K}N\) 
interaction and aligns with existing experimental constraints. The results contribute 
to the ongoing efforts to determine the existence and properties of the \(K^-pp\) 
system and to enhance our understanding of few-body meson-baryon dynamics.

From an experimental point of view, the present calculations are relevant 
to a setup in which the outgoing neutron is detected. Using the measured 
neutron momentum and the known initial-state kinematics, the invariant mass 
of the $(\pi\Sigma p)$ subsystem can be reconstructed via the neutron 
missing-mass technique. Therefore, the quantity discussed throughout this 
work as the neutron missing-mass spectrum is equivalent to the $\pi\Sigma p$ 
invariant-mass distribution. Although a fully exclusive measurement detecting 
all four final-state particles $(\pi,\Sigma,p,n)$ is, in principle, possible, 
the neutron missing-mass method provides a more practical and experimentally 
feasible approach.

The paper is organized as follows. Section~\ref{formula} introduces the formalism 
for the four-body $\bar{K}NNN$ system and outlines the transition amplitude for the 
$K^- + {}^3\mathrm{He}$ reaction. Section~\ref{result} discusses the two-body inputs 
and presents the calculated mass spectra. Finally, Sec.~\ref{conc} provides our 
conclusions.
\section{Formalism: four-body faddeev equations}
\label{formula}
One can investigate the properties of the $\bar{K}NNN$ system using the Faddeev 
Alt–Grassberger–Sandhas (AGS) formalism, focusing specifically on the state with 
total angular momentum $J = \tfrac{1}{2}$ and isospin $I = 0$. By treating the 
three nucleons as identical particles and simplifying the analysis by neglecting 
their intrinsic spin and isospin degrees of freedom, the system can be described 
in terms of three distinct partitions:
\begin{enumerate}
    \item Partition $\sigma = 1$: A configuration consisting of an antikaon ($\bar{K}$) 
    and a three-nucleon cluster ($NNN$), denoted as $[\bar{K} + (NNN)]$.
    \item Partition $\sigma = 2$: A single nucleon ($N$) coupled with a $\bar{K}NN$ 
    subsystem, represented as $[N + (\bar{K}NN)]$.
    \item Partition $\sigma = 3$: A system composed of a $\bar{K}N$ pair and an $NN$ pair, 
    described as $[(\bar{K}N) + (NN)]$.
\end{enumerate}

Before we proceed to present the four-body formalism, by systematically 
addressing each partition and employing the Faddeev-AGS formalism, we 
aim to shed light on the intricate interactions within the $\bar{K}NNN$ 
system and identify possible quasi-bound states that may emerge from these 
interactions. In partitions $\sigma=1$ and $\sigma=2$, the quasi-particles 
are three-body systems. To obtain the observables related to these subsystems, 
one should solve the three-body Faddeev equations. For a system including 
three particles $i$, $j$, and $k$, the three-body Faddeev equations in the 
AGS form can be expressed as~\cite{alt}:

\begin{equation}
\mathcal{K}_{ij,I_{i} I_{j}}^{\sigma}=
\mathcal{M}_{ij,I_{i} I_{j}}^{\sigma}+\sum_{k,I_{k}}
\mathcal{M}_{ik,I_i I_k}^{\sigma}
\tau_{kI_k}^{\sigma}\mathcal{K}_{kj,I_k I_j}^{\sigma}, 
\,\,\sigma=1,2
\label{eq1}
\end{equation}
where $\mathcal{K}_{ij,I_{i} I_{j}}^{\sigma}$ represents the transition 
operators, $\mathcal{M}_{ij,I_{i} I_{j}}^{\sigma}$ are the driving terms, 
and $\tau_{kI_k}^{\sigma}$ is an energy-dependent part of a separable 
two-body $T$-matrices within the three-body system. In the Faddeev 
equations~\ref{eq1}, the isospin quantum number of the interacting particle 
pair \((j,k)\), with particle \(i\) as the spectator, is denoted by \(I_i\). 
Solving these equations provides insight into the dynamics and possible 
bound or quasi-bound states of the three-body subsystems, such as 
$^{3}\mathrm{He}$, $\bar{K}(NN)_{s=0}$ and $\bar{K}(NN)_{s=1}$, which are 
essential for understanding the structure and reaction dynamics of the full 
four-body $\bar{K}NNN$ system.

The Faddeev equations in (\ref{eq1}) can give complete description for 
three-nucleon subsystem. For $\bar{K}(NN)_{s=0,1}$ since the \( \bar{K}N \) 
is coupled with $\pi\Sigma$ channel, one should consider the $\pi\Sigma{N}$ 
channel into the calculations. Therefore, in Faddeev equations (\ref{eq1}) 
plus the Faddeev partition indices $i,j,k=1,2,3$, one should modify the 
Faddeev equations to take the $\bar{K}N-\pi\Sigma$ coupling directly into 
account. Thus, in addition to the pair of Fadeev indices, each transition 
operator or driving term in Eq. (\ref{eq1}) should also carry a pair of 
particle channel indices, of type $\alpha,\beta,\gamma\,.\,.\,.$, each one 
having three possibilities, namely \( \bar{K}N_{1}N_{2}\), \(\pi\Sigma{N}_{2}\) 
or \(\pi{N}_{1}\Sigma\). See \cite{k4,k5,revei} for details.

In our three- and four-body calculations, the so-called exact optical 
potential method~\cite{shev3} was employed to simplify the Faddeev equations. 
This method avoids the complexity of solving the full three- and four-body 
equations, which are computationally demanding and may suffer from convergence 
difficulties. It provides a substantial simplification while retaining a high 
level of accuracy, making it a reliable approximation for treating coupled 
channels. Consequently, when solving the Faddeev equations (Eq.~(\ref{eq1})) 
for the $\bar{K}NN-\pi\Sigma N$ system, the $\pi\Sigma N$ channel was not 
included explicitly. The unnecessary particle channel indices have been 
omitted. 

Since the total isospin of the $\bar{K}NN$ system is $I=1/2$, therefore, 
depending on the spin of the two baryons, we should treat $K^{-}pp$ or 
$K^{-}d$ system. The total isospin of the interacting particles determines 
their total spin quantum number. Consequently, spin indices of the baryons 
do not appear explicitly, and the operators are labeled solely by isospin 
indices. In the $K^{-}pp$ system the spin component is antisymmetric, so all 
operators in isospin base should be symmetric and in the case of $K^{-}d$, 
all operators in isospin base should be antisymmetric.

To solve the three-body equations, it is beneficial to represent the 
three-body amplitudes and driving terms in a separable form. This approach 
transforms the equations (\ref{eq1}) into a homogeneous form, facilitating 
their solution. In this work, we employ the EDPE (Energy-Dependent Pole 
Expansion) method , as detailed in Refs.~\cite{nadro,naka,fix}. The separable 
representation of the Faddeev transition amplitudes is expressed as:
\begin{equation}
\mathcal{K}_{ij,I_i I_j}^{\sigma}(q,q';\epsilon_{\sigma})=
\sum_{\mu{,}\nu}^{N_{r}} 
u^{\sigma}_{\mu,iI_{i}}(q,\epsilon_{\sigma}) 
\theta^{\sigma;I_{i}I_{j}}_{ij;\mu\nu}(\epsilon_{\sigma}) 
u^{\sigma}_{\nu,jI_{j}}(q',\epsilon_{\sigma}),
\label{eq2}
\end{equation} 
where $u^{\sigma}_{\mu,iI_{i}}$ and $u^{\sigma}_{\nu,jI_{j}}$ are the form factors 
associated with the interacting subsystems, and 
$\theta^{\sigma;I_{i}I_{j}}_{ij;\mu\nu}(\epsilon_{\sigma})$ are the elements of the 
coupling matrix that encapsulate the interaction dynamics between different partitions. 
The variables $q$ and $q'$ represent the momenta of the spectator particle in channels 
$i$ and $j$ of the three-body subsystem, respectively.

The eigen functions $u^{\sigma}_{\mu,iI_{i}}(q,\epsilon_{\sigma})$ in Eq. (\ref{eq2}) 
can be defined by solving the homogeneous Faddeev equations  
\begin{equation}
\begin{split}
& u^{\sigma}_{\mu,iI_{i}}(q,\epsilon_{\sigma})=
\frac{1}{\lambda^{\sigma}_{\mu}}
\sum\limits_{jI_{j}}\int 
\mathcal{M}^{\sigma}_{ij,I_{i}I_{j}}(q,q';\epsilon_{\sigma})\, \\
& \hspace{2cm}\times \tau^{\sigma}_{jI_{j}} 
\big(\epsilon_{\sigma}-\frac{{q'}^2}{2M^{\sigma}_{j}}\big) 
u^{\sigma}_{\mu,jI_{j}}(q',\epsilon_{\sigma})d\vec{q}^{\prime},
\end{split}
\label{eq3}
\end{equation}
by solving Eq.~(\ref{eq3}), one can determine the possible binding energy 
\( B_{\sigma} \) of the three-body system, along with the corresponding 
form factors \( u^{\sigma}_{\mu,iI_{i}}(q,B_\sigma) \) and eigenvalues 
\( \lambda^{\sigma}_{\mu} \) at the energy \( \epsilon_{\sigma}=B_{\sigma} \). 
The form factors can be normalized by the condition
\begin{equation}
\begin{split}
& \sum_{i=1}^{3} \int u^{\sigma}_{\mu,iI_{i}}(q,B_\sigma) 
\tau^{\sigma}_{iI_{i}} 
\big(q,B_\sigma\big) 
u^{\sigma}_{\nu,iI_{i}}(q,B_\sigma)d\vec{q} \\ & 
\hspace{6cm} =-\delta_{\mu\nu}.
\end{split}
\label{eq4}
\end{equation}

In Eq.~(\ref{eq3}), the form factors are defined at a fixed energy 
\( \epsilon_{\sigma} = B_{\sigma} \), corresponding to the binding energy 
of the three-body system. To extend the applicability of the eigenfunctions 
\( u^{\sigma}_{\mu,iI_{i}} \) across the entire energy and momentum 
spectrum, we perform an extrapolation by
\begin{equation}
\begin{split}
& u^{\sigma}_{\mu,iI_{i}}(q,\epsilon_{\sigma})=
\frac{1}{\lambda^{\sigma}_{\mu}}
\sum\limits_{jI_{j};\beta}\int
\mathcal{M}^{\sigma}_{ij,I_{i}I_{j}}(q,q';\epsilon_{\sigma})\, \\
& \hspace{2cm}\times \tau^{\sigma}_{jI_{j}} 
\big(q',B_\sigma\big) 
u^{\sigma}_{\mu,jI_{j}}(q',B_\sigma)
d\vec{q}^{\prime}.
\end{split}
\label{eq5}
\end{equation}

After determining the eigenfunctions \( u^{\sigma}_{\mu,iI_{i}}(q,\epsilon_{\sigma}) \), 
one can define the effective EDPE propagators \( \theta(\epsilon_{\sigma}) \) 
in Eq.~(\ref{eq2}) by
\begin{equation}
\begin{split}
& \big((\theta^{\sigma}(\epsilon_{\sigma}))^{-1}\big)_{\mu\nu}
=\sum_{iI_{i}}\int
\big[u^{\sigma}_{\mu,iI_{i}}(q,B_\sigma)
\tau^{\sigma}_{iI_{i}} \big(q,B_\sigma\big) \\
& \hspace{1.cm} -u^{\sigma}_{\mu,iI_{i}}(q,\epsilon_{\sigma})
\tau^{\sigma}_{iI_{i}}\big({q,\epsilon_{\sigma}}\big)
\big] u^{\sigma}_{\nu,iI_{i}}(q,\epsilon_{\sigma})d\vec{q}.
\end{split}
\label{eq6}
\end{equation}

Based on Eq.~(\ref{eq6}), the Faddeev indices ($i,j$ and $k$) and isospin 
indices ($I_{i}$ and $I_{j}$) of the $\theta^{\sigma}(\epsilon_{\sigma})$
-functions are unnecessary and could be omitted. Therefore, we have
\begin{equation}
\begin{split}
\theta^{\sigma;I_{i}I_{j}}_{ij;\mu\nu}(\epsilon_{\sigma}) = 
\theta^{\sigma}_{\mu\nu}(\epsilon_{\sigma}).
\end{split}
\label{eq12}
\end{equation}

In addition to solving the three-body subsystems, it is essential to address 
the Faddeev equations for configurations involving two independent pairs of 
interacting particles, referred to as two quasi-particles. Specifically, we 
consider the $(\bar{K}N)(NN)_{s=0,1}$ subsystem. To analyze this state, we 
define the corresponding eigenfunctions. The pertinent equations are given by: 
\begin{equation}
\begin{split}
& \mathcal{Y}^{\sigma}_{ij;I_{i}I_{j}}=
\mathcal{W}^{\sigma}_{ij;I_{i}I_{j}}+
\mathcal{W}^{\sigma}_{ij;I_{i}I_{j}}
\tau^{\sigma}_{jI_{j}}\mathcal{Y}^{\sigma}_{jj;I_{j}I_{j}}, \\
& \mathcal{Y}^{\sigma}_{jj;I_{j}I_{j}}=
\mathcal{W}^{\sigma}_{ji;I_{j}I_{i}}
\tau^{\sigma}_{iI_{i}}
\mathcal{Y}^{\sigma}_{ij;I_{i}I_{j}}, \,\,\,\,\, \sigma=3,
\end{split}
\label{eq7}
\end{equation}
where the operators $\mathcal{Y}^{\sigma}_{ij;I_{i}I_{j}}$ are the Faddeev 
amplitudes which describe two independent pairs of interacting particles 
and the operator $\mathcal{W}^{\sigma}_{ij;I_{i}I_{j}}$ is the effective 
potential. In the same way as it was in the [3+1] subsystems, the 
eigenfunctions $u_{\mu;iI_{i}}^{\sigma}$ can be defined by solving the 
homogeneous Faddeev equations
\begin{equation}
u_{\mu;iI_{i}}^{\sigma}(q,\epsilon_{\sigma})=
\frac{1}{\lambda^{\sigma}_{\mu}} 
\mathcal{W}^{\sigma}_{ij;I_{i}I_{j}}(q,q';\epsilon_{\sigma})
\tau^{\sigma}_{jI_{j}}(\epsilon_{\sigma}) 
u_{\mu;jI_{j}}^{\sigma}(q',\epsilon_{\sigma}).
\label{eq8}
\end{equation}

Now that we have found the solution for the subsystems, we can go to solve the 
Faddeev equations for $\bar{K}NNN$ four-body system. Transition amplitudes 
between all possible partitions of the $\bar{K}NNN$ are given by~\cite{s19-2,shev5}.
\begin{equation}
\begin{split}
& \mathcal{A}^{I_{i}I_{j},\mu\nu}_{\sigma(i)\rho(j)}=
\mathcal{R}^{I_{i}I_{j},\mu\nu}_{\sigma(i)\rho(j)}+
\sum_{\omega;kl;\lambda\kappa}\sum_{I_{k},I_{l}}
\mathcal{R}^{I_{i}I_{k},\mu\lambda}_{\sigma(i)\omega(k)} \\
& \,\hspace{4.6cm}\times\theta^{\omega;I_{k}I_{l}}_{kl;\lambda\kappa} \, 
\mathcal{A}^{I_{l}I_{j},\kappa\nu}_{\omega(l)\rho(j)},
\end{split}
\label{eq9}
\end{equation}
where $\mathcal{A}^{I_{i}I_{j},\mu\nu}_{\sigma(i)\rho(j)}$ is the Faddeev 
transition amplitude from $\rho$ to $\sigma$ partition. To identify the 
spectator particle or interacting particles in each two- and three-body 
subsystem, we utilized the indices $i, j$ and $k$. The indices $\mu$ and $\nu$ 
refer to the separable expansion of the subamplitudes (equation~\ref{eq2}). 
The operators $\mathcal{R}^{I_{i}I_{j},\mu\nu}_{\sigma(i)\rho(j)}$ represent 
the driving terms, which describe the effective potential realized by the 
exchanged particle or quasi-particle between the channels $\sigma$ and $\rho$
\begin{equation}
\begin{split}
& \mathcal{R}^{I_{i}I_{j},\mu\nu}_{\sigma(i)\rho(j)}(p,p',E)=
\frac{\Omega^{I_{i}I_{j}}_{\sigma(i)\rho(j)}}{2}
\int^{+1}_{-1}d(\hat{p}\cdotp\hat{p}') \\
& \hspace{1.8cm} \times 
{u}^{\sigma}_{\mu,iI_{i}}(q,\epsilon_{\sigma})
\tau^{\sigma}_{iI_{i}}(p,p';z) 
u^{\rho}_{\nu,jI_{j}}(q',\epsilon_{\rho}).
\end{split}
\label{eq10}
\end{equation}

The symbols $\Omega^{I_{i}I_{j}}_{\sigma(i)\rho(j)}$ denote the 
Clebsch–Gordan coefficients. For a detailed explanation of the parameters 
in Eq.~(\ref{eq10}), the reader is referred to Refs.~\cite{s19-2,sa5}. 
The functions ${u}^{\sigma}_{\mu,iI_{i}}$ are the form factors that arise 
in the separable representation of the sub-amplitudes, as defined in 
Eqs.~(\ref{eq3}) and (\ref{eq8}). 

Looking for a scattering amplitude needs solution of the inhomogeneous system 
of equations (\ref{eq9}), which can be written in a matrix form
\begin{equation}
\mathcal{A}=\mathcal{R}+\mathcal{K}\mathcal{A},
\label{eq11}
\end{equation}
where the operator $\mathcal{K}$ is the kernel of the four-body Faddeev equations. 
The four-body scattering amplitude can be defined by
\begin{equation}
\mathcal{A}=(I-\mathcal{K})^{-1}\mathcal{R}.
\label{eq12}
\end{equation}

In this work, we focus on the \( K^{-} + \mathrm{^{3}He} \) reaction as a 
potential avenue for observing the $K^{-}pp$ bound state. We analyze one 
specific reaction channel:
\begin{equation}
K^{-} + \mathrm{^{3}He} \rightarrow n + [\bar{K}NN]_{s=0} \rightarrow 
n + \pi\Sigma p.
\nonumber
\end{equation}

In this scenario, the intermediate state consists of a neutron and a \(\bar{K}NN\) 
three-body system with total spin \(s = 0\). By analyzing the neutron missing mass 
spectrum in this specific reaction channel, we aim to identify a potential signal 
corresponding to the formation of the \(K^{-}pp\) bound state. A distinct peak in 
the missing mass spectrum would provide evidence supporting the existence of this 
deeply bound state.

The transition from the initial \( K^{-} + \mathrm{^{3}He} \) state to the 
final \( n + \pi\Sigma p \) channel involves a series of intricate processes, 
as illustrated in Figure~\ref{fig0}. Initially, the system undergoes a 
transformation from \( K^{-} + \mathrm{^{3}He} \) to an intermediate state 
characterized by \( N + (\bar{K}NN)_{s=0} \). This stage represents a complex 
four-body interaction, encompassing an antikaon (\( K^{-} \)) and a helium-3 
nucleus (\( \mathrm{^{3}He} \)). To accurately determine the transition 
amplitude for this process, it is essential to solve the inhomogeneous 
four-body Faddeev equations (\ref{eq9}). 

Subsequently, the intermediate \( (\bar{K}NN)_{s=0} \) system, in the presence 
of a spectator nucleon (\( N \)), decays into the \( \pi\Sigma N \) channel. 
This decay process is influenced by the isospin state (\( I=0 \) and \( I=1 \)) 
of the \( (\bar{K}N) \) system, which dictates the possible decay pathways and 
the resulting particle configurations. In summary, the transition from 
\( K^{-} + \mathrm{^{3}He} \) to \( n + \pi\Sigma p \) encompasses a sequence 
of complex interactions, including the formation of an intermediate four-body 
state and its subsequent decay. Employing the Faddeev formalism provides a 
robust framework for analyzing these processes, offering valuable insights into 
the underlying physics of few-body systems and the interactions governing them. 
Therefore, the scattering amplitude ($\mathcal{T}$) for the 
$K^{-} + \mathrm{^{3}He} \rightarrow  n + (\pi\Sigma p)$ reaction channel can 
be defined by~\cite{kha}
\begin{equation}
\begin{split}
& \mathcal{T}= 
 \sum_{I_{NN}=0,1}
 \sum_{I_{\bar{K}N}=0,1}
 \sum_{\lambda,\mu,\nu=1}^{N_{r}}
 \mathcal{A}^{I_{NN}I_{\bar{K}N},\mu\lambda}_{1(N)2(N)} 
 (P_{K^{-}},p_{n};E) \\
 & \,\hspace{1.0cm} \times \theta^{\rho=2}_{\lambda\nu} (p_{n},E)
\, u^{\rho=2}_{\nu,pI_{p}}(p_{n},q_{p};E) \\
& \,\hspace{1.0cm} \times 
\tau^{I=0,1}_{\bar{K}N\rightarrow\pi\Sigma}(p_{n},q_{p};E)
\,\, g^{I=0,1}_{\pi\Sigma}(p_{n},q_{p};E),
\end{split}
\label{eq13}
\end{equation}
where $p_{n}$ denotes the momentum of the spectator neutron, $q_{p}$ 
represents the spectator proton momentum in the $K^{-}pp$ three-body 
subsystem when the proton acts as the spectator particle. In Eq.~\ref{eq13}, 
the functions $u^{\rho}_{\nu,iI_{i}}$ and $\theta^{\rho}_{\lambda\nu}$ 
were obtained by solving the one-channel Faddeev AGS equations. 
The total cross section in center of mass coordinates can be calculated by
\begin{equation}
\begin{split}
& \sigma=(4\pi)^{3} \int p^{2}_{n}dp_{n}\int q_{p}^{2}dq_{p}\int 
k_{\pi\Sigma}^{2}dk_{\pi\Sigma} \delta(E_{f}-E_{i}) \\
& \hspace{2cm}\times \frac{(2\pi)^{4}}{|\vec{v}_{K^{-}}-\vec{v}_{\mathrm{^{3}He}}|} 
\,\,|\mathcal{T}|^{2},
\end{split}
\label{eq133}
\end{equation}
where $\mathcal{T}$ is the scattering amplitude calculated in Eq.\ref{eq13} and 
the energies $E_{i}$ and $E_{f}$ are given by 
\begin{equation}
\begin{split}
& E_{i}=\frac{P_{K^{-}}^{2}}{2m_{K^{-}}}+\frac{P_{K^{-}}^{2}}{2m_{\mathrm{^{3}He}}}
+\Delta{m}c^{2}-B_{\mathrm{^{3}He}} \\
& E_{f}=\frac{k_{\pi\Sigma}^{2}}{2\mu_{\pi\Sigma}}+
\frac{p_{n}^{2}(m_{n}+m_{p}+m_{\pi}+m_{\Sigma})} {2m_{n}(m_{p}+m_{\pi}+m_{\Sigma})} \\
& \hspace{0.4cm}+\frac{q_{p}^{2}(m_{p}+m_{\pi}+m_{\Sigma})} {2m_{p}(m_{\pi}+m_{\Sigma})}.
\end{split}
\label{eq134}
\end{equation}

The $\delta$-function can be written in terms of $k_{\pi\Sigma}$ in the form
\begin{equation}
\delta(E_f - E_i) = \left|
\frac{1}{\left.\dfrac{dE_f}{dk_{\pi\Sigma}}\right|_{k_{\pi\Sigma}=k_{\pi\Sigma}^0}}
\right| \, \delta(k_{\pi\Sigma} - k_{\pi\Sigma}^0),
\end{equation}
where the momentum $k^{0}_{\pi\Sigma}$ is defined by
\begin{equation}
\begin{split}
k_{\pi\Sigma}^0 = \sqrt{2\mu_{\pi\Sigma}}
\Bigg[ E_i
- \frac{p_n^2 (m_n + m_p + m_\pi + m_\Sigma)}
     {2 m_n (m_p + m_\pi + m_\Sigma)} \\
\quad
- \frac{q_p^2 (m_p + m_\pi + m_\Sigma)}
       {2 m_p (m_\pi + m_\Sigma)}
\Bigg]^{1/2}.
\end{split}
\end{equation}

Therefore, the neutron missing mass can be writen in the form
\begin{equation}
\begin{split}
& d\sigma/dE_{n} =(4\pi)^{3} \int q_{p}^{2}dq_{p} 
\frac{(2\pi)^{4} m_{K^{-}}m_{\mathrm{^{3}He}}}{ P_{K^{-}} 
(m_{K^{-}}+m_{\mathrm{^{3}He}})} \\
& \hspace{2cm}\times k^{0}_{\pi\Sigma}\, \mu_{\pi\Sigma}\, p_{n} \,
|\mathcal{T}|^{2},
\end{split}
\label{eq135}
\end{equation}
where $E_n$ denote the energy of the outgoing neutron.

Although the differential cross section is formally calculated as a function 
of the neutron energy, $d\sigma/dE_n$, the results are presented in terms of 
the invariant mass of the $\pi\Sigma p$ subsystem. This choice reflects the 
experimental relevance of the $\pi\Sigma p$ invariant-mass spectrum and its 
direct connection to the $K^-pp$ and $\Lambda(1405)$ signals.

The two variables are related through energy conservation in the center-of-mass 
frame. For a fixed initial energy $E_{\mathrm{tot}}$, the invariant mass of the 
$\pi\Sigma p$ system is given by
\begin{equation}
\begin{split}
& M_{\pi\Sigma p} = m_{\pi}+m_{\Sigma}+m_{p}+(E_{i}-E_{n}) \\
& \hspace{0.9cm}=m_{\pi}+m_{\Sigma}+m_{p}+(E_{i}-\frac{p^{2}_{n}}{2m_{n}}).
\end{split}
\label{eq136}
\end{equation}

Thus, the neutron missing-mass spectrum and the $\pi\Sigma p$ invariant-mass 
spectrum represent equivalent descriptions of the same physical observable, 
related by a change of kinematical variables. In the present work, we use 
$M_{\pi\Sigma p}$ as the horizontal axis in the figures for clarity.
\section{Results and Discussions}
\label{result}
In this paper, we consider all two-body interactions to have zero orbital 
angular momentum. The primary interactions examined are between antikaon
-nucleon ($\bar{K}N$) and nucleon-nucleon ($NN$) pairs. The $\bar{K}N$ 
subsystem predominantly couples with the pion-Sigma ($\pi\Sigma$) channel 
in isospin states $I=0$ and with $\pi\Sigma$ and $\pi\Lambda$ channels in 
$I=1$. For the $\bar{K}N$ interaction, we utilize separable potential in 
the momentum representation, defined as:
\begin{equation}
V_{I}^{\alpha\beta}(k^{\alpha},k^{\beta})
=g_{I}^{\alpha}(k^{\alpha})\,
\lambda_{I}^{\alpha\beta}\,
g_{I}^{\beta}(k^{\beta}),
\label{eq14}
\end{equation}
where $g_{I}^{\alpha}(k^{\alpha})$ is the form factor of the interacting two-body 
subsystem, with relative momentum $k^{\alpha}$ and isospin $I$ and 
$\lambda_{I}^{\alpha\beta}$ is the strength parameter of the interaction. To take 
the $\bar{K}N-\pi\Sigma$ coupling directly into account, the potential is further 
labeled with the $\alpha$ values. The two-body $T$-matrices in separable form can 
be given by
\begin{equation}
\begin{split}
T_{I}^{\alpha\beta}(k^{\alpha},k^{\beta};z)
=g_{I}^{\alpha}(k^{\alpha})\,
\tau_{I}^{\alpha\beta}(z)\,g_{I}^{\beta}(k^{\beta}),
\end{split}
\label{eq15}
\end{equation}
where $z$ is the total energy of the two-body subsystem and the quantity 
$\tau_{i,I_{i}}^{\alpha\beta}(z)$ is an energy-dependent part of a separable 
two-body $T$-matrix.

To describe the coupled-channel \( \bar{K}N\text{--}\pi\Sigma \) interaction, 
which plays a crucial role in both our three- and four-body calculations, 
we employ two different phenomenological potentials and one chiral potential. 
These potentials are constructed to reproduce either a one-pole or a two-pole 
structure of the \( \Lambda(1405) \) resonance. The parameters of the 
phenomenological potentials, denoted as \(\mathrm{SIDD1}\) and \(\mathrm{SIDD2}\), 
are taken from Ref.~\cite{shev4}, while those of the energy-dependent chiral 
potential are given in Ref.~\cite{k10}. The corresponding pole positions 
for each potential model are summarized in Table~\ref{table1}. In addition 
to the $\bar{K}N$ pole position, we have also calculated the $K^-pp$ pole 
position using the one-channel AGS equations.

For modeling nucleon-nucleon ($NN$) interaction, we utilize the one-term PEST 
potential~\cite{pest}. This potential serves as a separable approximation to 
the Paris potential, focusing primarily on the attractive long-range component 
of the $NN$ interaction while omitting the short-range repulsive part. 
Consequently, this simplification leads to an overestimation of the binding 
energy in light nuclei, such as helium-3 ($B_{^{3}\mathrm{He}}=10.7$ MeV).
\begin{table*}[t]
\caption{
Model dependence of the \( \bar{K}N \) and \( \bar{K}(NN)_{s=0} \) pole 
positions (in MeV). Results are presented for the phenomenological SIDD1 
and SIDD2 potentials, as well as the chiral-based \( \bar{K}N\text{--}\pi\Sigma \) 
interaction. The SIDD1 model yields a single-pole structure, whereas the 
SIDD2 and chiral-based potentials reproduce the two-pole structure of the 
\( \Lambda(1405) \) resonance.}
\centering
\begin{tabular}{cccc}
\hline\hline\noalign{\smallskip}
&\,  $\mathrm{SIDD1}$ \, & \, $\mathrm{SIDD2}$ \, & \, $\mathrm{Chiral}$  \\
\noalign{\smallskip}\hline
\noalign{\smallskip}
$\bar{K}N$ first pole \, & \, $1428.1-i46.6$~\cite{shev4} \, 
& \, $1418.1-i56.9$~\cite{shev4} \, & \, $1420.6-i20.3$~\cite{k10} \\
$\bar{K}N$ second pole \, & \, $-$ \, & \, $1382.0-i104.2$~\cite{shev4} \, 
& \, $1343.0-i72.5$~\cite{k10} \\
\hline\noalign{\smallskip}
$K^{-}pp$ pole position \, & \, $2325-i34$ \, & \, $2325-i24$ \, & \, $2346-i22$ \\
\noalign{\smallskip}
\hline\hline
\end{tabular}
\label{table1} 
\end{table*}

Since the kernel of AGS equations has the standard moving singularities that 
are caused by the opened channels and are encountered in any three and four-body 
breakup problem, we have followed the same procedure implemented in Refs.~\cite{p1,p2}. 
In few-body scattering calculations, particularly those employing the Faddeev 
or Alt–Grassberger–Sandhas (AGS) formalism, the kernel integrals often contain 
singular terms of the form
\begin{equation}
\frac{1}{E - E_p + i\varepsilon},
\end{equation}
which become problematic when the total energy \(E\) approaches the pole position 
\(E_p\). Direct numerical integration in this region is unstable due to the 
divergence induced by the vanishing denominator. To handle this issue, the 
\emph{point method} is frequently applied~\cite{p1,p2}.

The core idea of the point method is to avoid evaluating the integral exactly 
at the singular energy. Instead, one computes the scattering amplitude (or any 
required quantity) at several nearby complex energies
\begin{equation}
E_j = E + i\varepsilon_j, \qquad \varepsilon_j > 0,
\end{equation}
where \(\varepsilon_j\) are small positive shifts that move the integration 
contour off the real axis. At each shifted energy \(E_j\), the integrals are 
well-defined and can be evaluated numerically without encountering 
divergences:
\begin{equation}
X(E_j) = \int_{0}^{\infty} \frac{f(p)}{E_j - p^2} \, dp.
\end{equation}

Once the values \(X(E_j)\) have been obtained for several \(\varepsilon_j\), 
one assumes analyticity of \(X(E)\) in the vicinity of the real axis and performs 
an analytic continuation:
\begin{equation}
X(E) = \lim_{\varepsilon \to 0^{+}} X(E + i\varepsilon).
\end{equation}

In practice, this continuation is achieved by fitting \(X(E_j)\) as a function 
of \(\varepsilon_j\) using either a polynomial expansion or a Padé approximant. 
This approach avoids the need for explicit principal-value integration or complex 
contour deformation, while maintaining numerical stability. It is particularly 
advantageous for above-threshold energies, where singularities are unavoidable and 
computational efficiency is essential.

\begin{figure}[h]
\centering
\includegraphics[width=8.6cm]{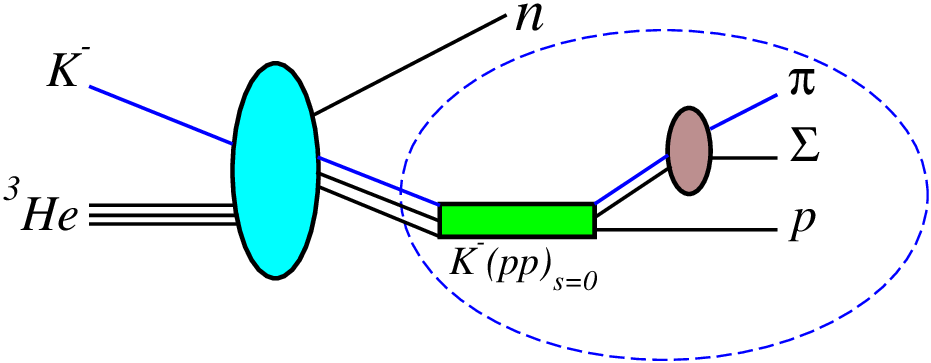} 
\caption{(Color online) Schematic representation of the transition 
from \( K^{-} + {}^{3}\mathrm{He} \) to \( n + (\pi\Sigma p) \) via 
the intermediate state \( N + (\bar{K}NN)_{s=0} \). The large solid 
ellipse denotes the four-body dynamics of the initial system, while 
the rectangular box represents the three-body dynamics associated with 
the formation and decay of the \( \bar{K}NN \) cluster.}
\label{fig1}
\end{figure}

\begin{figure}[h]
\hspace{-0.3cm}
\includegraphics[width=8.8cm]{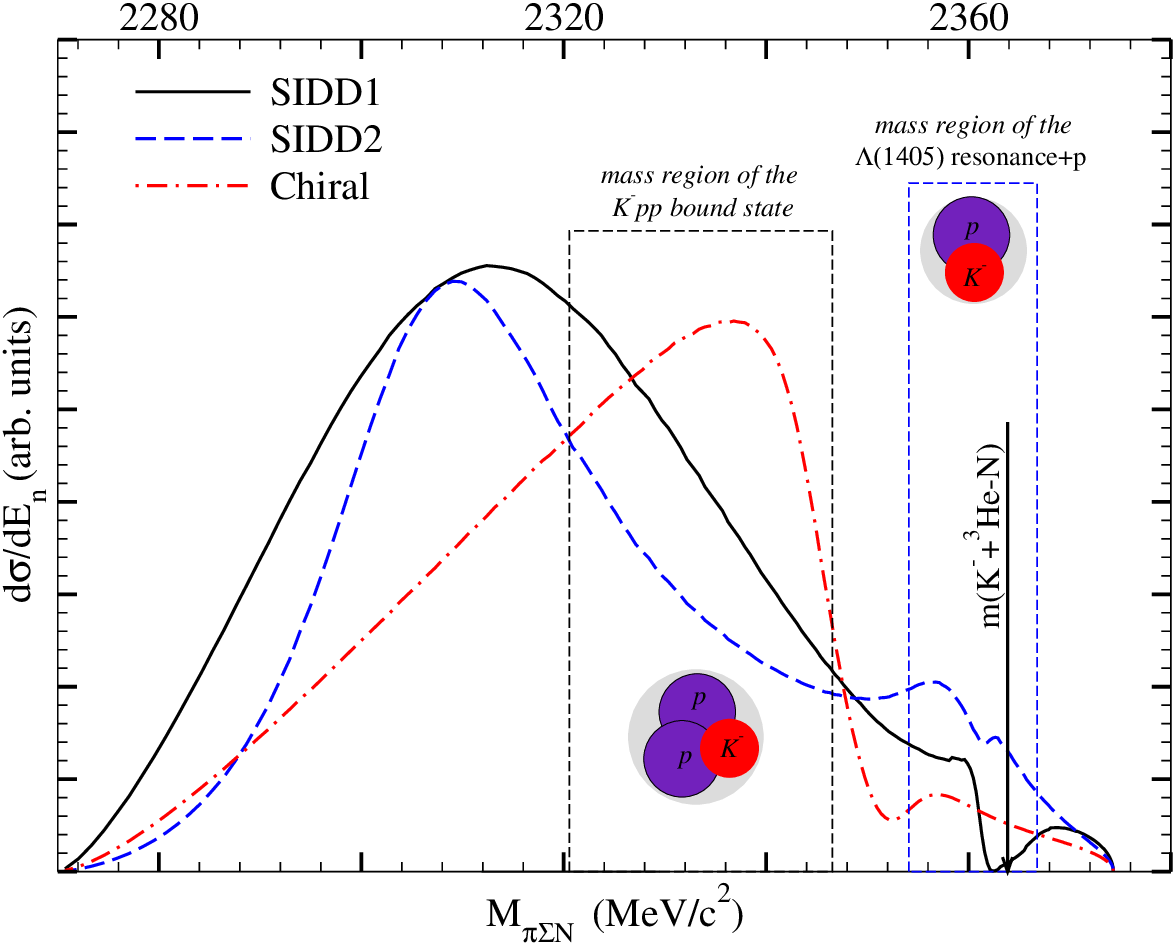} \\
\vspace{-0.5cm}
\caption{(Color online) 
The neutron missing mass spectrum is presented. In this analysis, the 
full four-body Faddeev amplitude is not included; instead, only the 
reaction mechanisms within the region within the dashed ellipse in 
Fig.~\ref{fig1} are considered. A clear peak associated with the \( K^-pp \) 
bound state is visible for all interaction models examined. In addition 
to the \( K^-pp \) signal, an enhancement attributed to the \( \Lambda(1405) \) 
resonance appears in the spectrum, particularly when using the SIDD2 and 
chiral potentials. In contrast, for the SIDD1 potential, the \( \Lambda(1405) \) 
pole lies above the threshold, which is reflected in the corresponding 
spectral distribution.
}
\label{fig2}
\end{figure}

Using the transition amplitude defined in Eq.~(\ref{eq13}), the missing mass 
of the spectator nucleon (\( d\sigma/dE_{n} \)) was calculated. As a first step, 
the four-body transition amplitude in Eq.~(\ref{eq13}) was turned off by setting 
\(\mathcal{A}^{I_{i}I_{j},\mu\lambda}_{\sigma(i)\rho(j)} = 1,\) and the neutron 
missing mass spectrum was computed. In this case, only the interactions within 
the dashed ellipse in Fig.~\ref{fig1} were included. The corresponding results 
are shown in Fig.~\ref{fig2}. 

The incoming antikaon momentum was set to 
\(
p_{K^{-}} = 100~\mathrm{MeV}/c,
\)
and the missing mass of the neutron was evaluated using three different 
\( \bar{K}N \) interaction models. In these calculations, the number of 
terms \( N_{r} \) in Eq.~(\ref{eq2}) was fixed to \( N_{r} = 2 \), which 
ensures sufficient numerical accuracy for practical purposes.

A pronounced peak associated with the \( K^-pp \) bound state appears 
for all interaction models. However, compared to the values reported 
in Table~\ref{table1}, the peak position is shifted by approximately 
10-15~MeV toward the \( \pi\Sigma p \) threshold, regardless of the 
chosen \( \bar{K}N \) potential. Additionally, a structure related to 
the \( \Lambda(1405) \) resonance emerges in the spectra obtained with 
the chiral-based and SIDD2 potentials. This feature appears as a shoulder 
on the main \( K^-pp \) peak. In contrast, for the SIDD1 potential, the 
\( \Lambda(1405) \) signal is strongly affected by threshold effects, 
since the \( \Lambda(1405) + p \) mass threshold lies very close to 
\(
M({}^{3}\mathrm{He}) + m(K^-) - M(n).
\)

Theoretical studies using Faddeev calculations have indicated that the formation 
of a bound \(K^-pp\) state could manifests as a distinct peak in the \(\pi\Sigma{N}\) 
invariant mass \cite{sh13,s19}. It was also shown in \cite{s19} that the 
\(\Lambda(1405)\) resonance manifests itself in the \(\pi\Sigma{N}\) mass spectrum. 
However, these calculations were three-body type approximations of the 
\(K^{-}+ \,^{3}\mathrm{He}\) interaction and the full dynamics of the four-body 
system and the effect of the spectator neutron were not included properly. The 
results presented in Fig.~\ref{fig1} are in good agreement with those reported 
in Ref.~\cite{s19}, where it was shown that, in addition to the \( K^-pp \) 
bound state signal, the signature of the \( \Lambda(1405) \) resonance appears 
in the \( \pi\Sigma p \) invariant mass spectrum.

\begin{figure}[h]
    \centering
    \hspace{-0.5cm}
    \includegraphics[width=8.8cm]{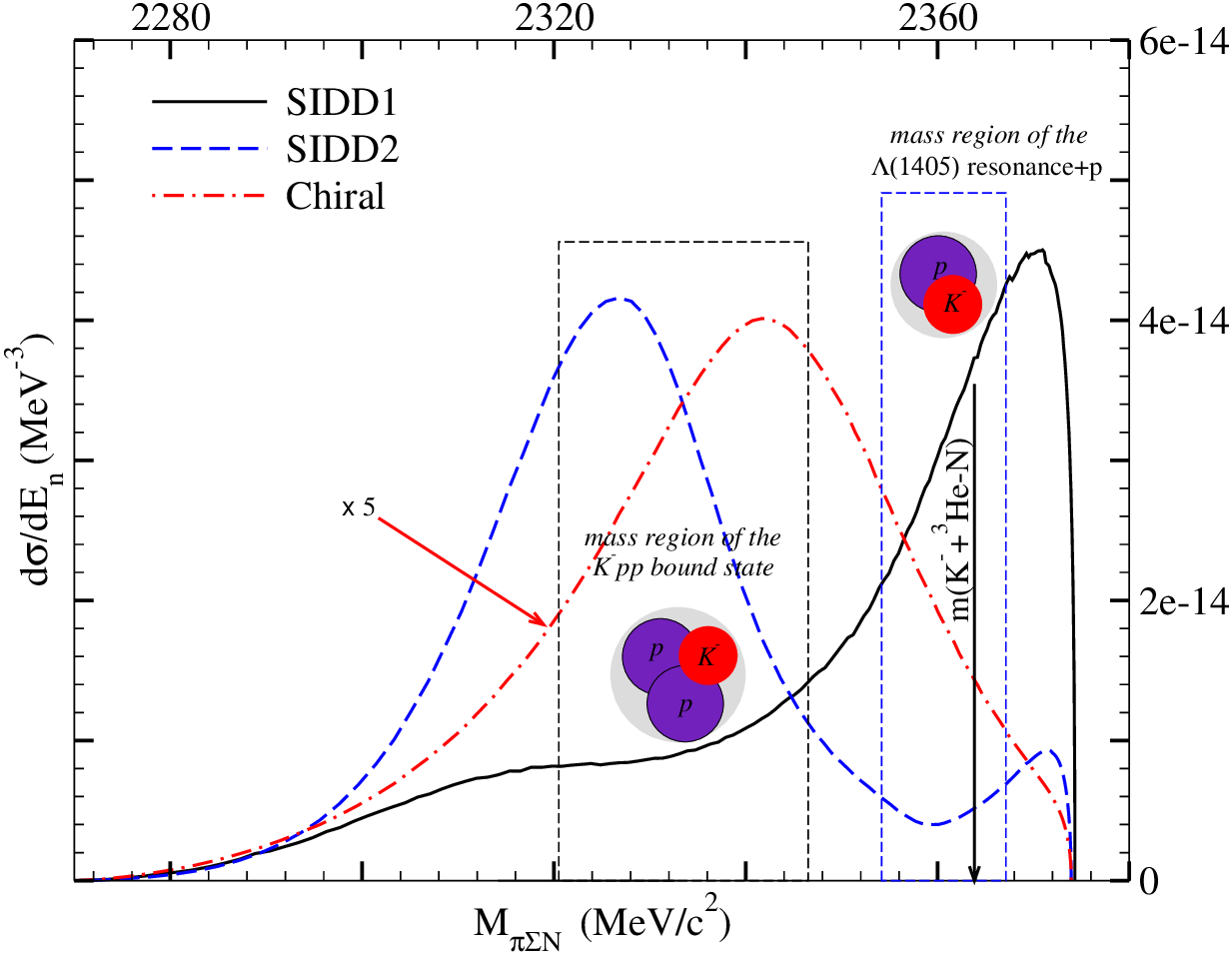}
    \vspace{-0.cm}
    \caption{Calculated neutron missing mass spectrum for the 
    $K^- + ^3$He $\rightarrow$ $n + (\pi\Sigma p)$ reaction 
    at a kaon momentum of 100~MeV/$c$. The explanations are the 
    same as in Fig.~\ref{fig2}. Here, the full four-body Faddeev 
    calculations was performed to detemine the neutron mass spectrum.}
    \label{khe3fig}
\end{figure}

In Fig.~\ref{khe3fig}, we performed a full four-body calculation of the reaction 
\( K^- + {}^3\text{He} \rightarrow (\pi \Sigma p) + n \) to analyze the spectral 
features in the final-state channels. Our approach incorporates the full dynamics 
of the antikaon interacting with the three-nucleon system, accounting for all 
relevant couplings and decay mechanisms leading to the \( (\pi \Sigma p) + n \) 
final state. The computed spectrum reveals a clear enhancement near the \( \bar{K}NN \) 
pole position threshold in the invariant mass distribution of the \( \pi \Sigma p \). 
The characteristics of the calculated spectrum particularly the peak location below 
the \( K^-pp \) threshold and its moderate width are not reproduced by non-resonant 
or purely kinematical mechanisms. This structure is consistent in both peak position 
and width with the pole energies reported in Table~\ref{table1}. This indicates that 
the observed enhancement is a genuine manifestation of strong \( \bar{K}N \) attraction 
in the isospin-zero channel, leading to a tightly bound few-body configuration. The 
results are qualitatively consistent with the enhancement observed by the E15 experiment 
and interpreted by Sekihara \textit{et al.} as a $K^-pp$-like state. However, at the 
lower kaon momentum used in our study, the peak is sharper and less distorted by 
background, supporting the hypothesis that low-energy $K^-$ beams are better suited for 
studying such quasi-bound states. This enhancement is attributed to the reduced momentum 
transfer and the resulting increased overlap between the kaon and the $^3$He system, 
facilitating the formation of a quasi-bound $\bar{K}NN$ configuration. Further exclusive 
measurements of the \( (\pi \Sigma p) \) spectrum, including angular correlations, will 
be valuable for constraining the binding energy and decay width of this exotic system 
and for advancing our understanding of antikaon-nuclear interactions.

A comparison of the \( \pi\Sigma p \) mass spectra obtained using the chiral-based 
potential (red dashed and dotted curve) and the SIDD2 potential (blue dashed curve) 
with those calculated within the three-body approximation (Fig.~\ref{fig1}) shows 
that a similar peak structure associated with the \( K^-pp \) bound state is observed 
in both cases. However, the peak corresponding to the \( \Lambda(1405) \) resonance 
does not clearly appear in the spectrum extracted from the four-body calculation, 
which is possibly due to the effects of four-body dynamics.

A clear difference is observed when comparing the three-body results in 
Fig.~\ref{fig2} with the full four-body calculations shown in Fig.~\ref{khe3fig}, 
most notably for the SIDD1 interaction. While the three-body dynamics 
exhibits a pronounced signal of the $K^-pp$ quasi-bound state together with 
the $\Lambda(1405)$ resonance, the inclusion of the spectator neutron and 
full four-body kinematics leads to a strong suppression of the $K^-pp$ 
signal and an enhancement of the $\pi\Sigma$ strength. This behavior can be 
attributed to the single-pole structure and stronger subthreshold attraction 
of the SIDD1 $\bar{K}N$ potential. In the four-body framework, recoil effects 
and additional off-shell dynamics redistribute the spectral strength toward 
the $\pi\Sigma$ channel, thereby favoring the $\Lambda(1405)$ contribution. 
In contrast, for the SIDD2 and chiral interactions, which exhibit a two-pole 
structure of the $\Lambda(1405)$, the transition from three- to four-body 
dynamics results in only moderate changes of the mass spectrum.

Recently, Shevchenko \cite{sh25} performed a coupled-channel three-body 
calculation including additional particle channels, leading to a less 
bound and broader $K^-pp$ quasi-bound state. A fully coupled-channel 
four-body treatment would be considerably more complex and is beyond 
the scope of the present work. Nevertheless, these results indicate that 
explicit particle-channel effects may further modify the properties of 
the $K^-pp$ system.

In this work, we present theoretical calculations of the reaction 
$K^- + {}^3\mathrm{He} \rightarrow n + (\pi\Sigma p)$, focusing on a low kaon 
momentum of $100$~MeV/$c$. The choice of a low-energy kaon is motivated by several 
advantages: near-threshold quasi-bound states, such as $K^- pp$, are more prominently 
formed and can be studied with greater clarity, while the reduced kinetic energy limits 
the phase space for background processes, making signals from the desired channels more 
distinct. Additionally, at lower energies, theoretical modeling becomes more controlled, 
as fewer reaction channels are open, allowing for a more precise treatment of the dynamics 
using few-body techniques. The low-momentum regime also aligns with the validity of the 
$\bar{K}N$ interaction models employed in this work (SIDD1, SIDD2, and Chiral potentials), 
which are effective primarily in the low-energy region. Moreover, our framework enables 
the calculation of the neutron missing-mass spectrum and absolute cross sections, 
providing direct guidance for experimental searches of the $K^- pp$ quasi-bound state. 
Although the J-PARC E15 experiment \cite{E1520} employs a higher kaon momentum of 
about $1$~GeV/$c$, the essential reaction mechanisms, including the formation of the 
$K^- pp$ quasi-bound state and contributions from the $\Lambda(1405)$ resonance, are 
qualitatively comparable. Hence, our results provide complementary insights into the 
reaction dynamics and serve as guidance for the interpretation of spectral shapes 
observed in E15 and future low-momentum kaon experiments.
\section{Conclusion}
\label{conc}
In conclusion, our four-body calculation of the 
$K^- + {}^3\mathrm{He} \to (\pi \Sigma p)+n$ reaction at an incoming kaon 
momentum of 100 MeV/c demonstrates that low-energy kaons are highly effective 
in probing $\bar{K}NN$ dynamics. The results show enhanced signal strength 
between the $\pi\Sigma{N}$ and $\bar{K}NN$ thresholds, supporting the 
interpretation of a $\bar{K}NN$ bound state. In comparison with E15 data and 
the theoretical framework of Sekihara \textit{et al.,} our findings suggest 
that low-momentum kaon beams offer significant advantages for identifying and 
studying such exotic states. Moreover, the comparison of different $\bar{K}N$ 
interaction models indicates that the spectral features are sensitive to the 
underlying coupled-channel dynamics, emphasizing the importance of a consistent 
treatment of the $\Lambda(1405)$ in few-body calculations. Our model, which 
fully accounts for the four-body nature of the initial system, offers a refined 
description compared to simpler few-body approaches. It accurately represents 
the nuclear environment and final-state dynamics, reinforcing the interpretation 
of the E15 signal as evidence for a \( \bar{K}NN \) bound state. Our findings 
suggest that future experiments optimized for low-energy kaon beams and improved 
neutron detection could provide clearer evidence for the existence and properties 
of $K^-pp$-like systems. The theoretical predictions made in this study pave the 
way for such investigations and contribute to the ongoing effort to understand 
the strong interaction in the strangeness sector.

\end{bibunit}
\end{document}